\documentclass[
aps,
twocolumn, % twocolumn onecolumn
showpacs,
prl, % prl preprint reprint,
superscriptaddress] %,floatfix]
{revtex4-1}
\usepackage{graphicx}
\usepackage{color}
\usepackage{ulem}
\usepackage{verbatim}
\usepackage{amsmath}
\usepackage{amssymb}
\usepackage{hyperref}

\hypersetup{
  colorlinks   = true, %Colours links instead of ugly boxes
  urlcolor     = blue, %Colour for external hyperlinks
  linkcolor    = blue, %Colour of internal links
  citecolor    = blue %Colour of citations
}

\begin{document}

%\title{Generation of time-bin entanglement between a photon and a collective atomic excitation}%
\title{Entanglement between a photonic time-bin qubit and a collective atomic spin excitation}%

% 42.50.Dv 	Quantum state engineering and measurements
% 03.67.Hk 	Quantum communication
% 32.80.Qk 	Coherent control of atomic interactions with photons

% 03.67.-a 	Quantum information
% 42.50.-p 	Quantum optics

% 03.65.Ud 	Entanglement and quantum nonlocality, Bell inequalities
% 03.67.Bg 	Entanglement production and manipulation
% 03.67.Mn 	Entanglement measures, witnesses, and other characterizations
% 42.50.Ex 	Optical implementations of quantum information processing and transfer
% 42.50.Md 	Optical transient phenomena: quantum beats, photon echo, free-induction decay, dephasings and revivals, optical nutation, and self-induced transparency

\author{Pau Farrera}
\email[Contact: ]{pau.farrera@icfo.eu}
\affiliation{ICFO-Institut de Ciencies Fotoniques, The Barcelona Institute of Science and Technology, 08860 Castelldefels (Barcelona), Spain}

\author{Georg Heinze}
\email[Contact: ]{georg.heinze@icfo.eu}
\affiliation{ICFO-Institut de Ciencies Fotoniques, The Barcelona Institute of Science and Technology, 08860 Castelldefels (Barcelona), Spain}

\author{Hugues de Riedmatten}
\homepage{http://qpsa.icfo.es}
\affiliation{ICFO-Institut de Ciencies Fotoniques, The Barcelona Institute of Science and Technology, 08860 Castelldefels (Barcelona), Spain}
\affiliation{ICREA-Instituci\'{o} Catalana de Recerca i Estudis Avan\c cats, 08015 Barcelona, Spain}%

%\date{\today}

\begin{abstract}
Entanglement between light and matter combines the advantage of long distance transmission  of photonic qubits with the storage and processing capabilities of atomic qubits. To distribute photonic states efficiently over long distances several schemes to encode qubits have been investigated -- time-bin encoding being particularly promising due to its robustness against decoherence in optical fibers. Here, we demonstrate the generation of entanglement between a photonic time-bin qubit and a single collective atomic spin excitation (spin-wave) in a cold atomic ensemble, followed by the mapping of the atomic qubit onto another photonic qubit. A magnetic field that induces a periodic dephasing and rephasing of the atomic excitation ensures the temporal distinguishability of the two time-bins and plays a central role in the entanglement generation. To analyse the generated quantum state, we use largely imbalanced Mach-Zehnder interferometers to perform projective measurements in different qubit bases and verify the entanglement by violating a CHSH Bell inequality.
\end{abstract}

\maketitle

Entangled states between light and matter play a central role for fundamental tests in quantum physics \cite{Blinov2004, Sherson2006, Hofmann2012, Hensen2015}. In addition they are an important resource for several emerging quantum technologies, such as quantum cryptography, quantum computation or remote sensing. Their main key feature is that they combine the advantages of ``flying" photonic states (that provide long distance transmission) with the ones of ``stationary" atomic states (that enable quantum state storage, synchronization and processing) \cite{Kimble2008, Sangouard2011}. These hybrid entangled states have been generated with a large variety of matter systems such as atomic gases \cite{Matsukevich2005, Riedmatten2006, Inoue2006,Chen2007,Dabrowski2017}, single atoms and ions \cite{Blinov2004, Volz2006, Wilk2007} or solid state systems \cite{Togan2010, Clausen2011, Saglamyurek2011}.

Several degrees of freedom can be used to encode the photonic component of the light-matter entangled state: polarization \cite{Blinov2004,Matsukevich2005,Riedmatten2006,Volz2006,Chen2007,Wilk2007,Togan2010,Hofmann2012}, orbital angular momentum \cite{Inoue2006}, spatial \cite{Dabrowski2017} or time-bin encoding \cite{Clausen2011, Saglamyurek2011,Hensen2015} being the most prominent examples. For long distance transmission, encoding in form of photonic time-bin qubits is favourable, since this approach is robust against decoherence in optical fibers. The good performance of time-bin photonic states has been shown in several works including long distance entanglement distribution or teleportation \cite{Marcikic2003, Marcikic2004, Valivarthi2016,Sun2016}. The direct generation of time-bin entanglement between a photonic qubit and a matter system (i.e. without the need of an external quantum light source) has so far been shown only in single emitters systems such as nitrogen vacancy centers \cite{Hensen2015} or quantum dots \cite{DeGreve2012}.

In this work, we demonstrate direct generation of entanglement between a photonic time-bin qubit and a collective atomic spin excitation (spin-wave) using an ensemble of laser cooled atoms. In contrast to former related experiments, in our system, the atomic state can be later mapped very efficiently into a single photon without the need of high finesse cavities \cite{Bimbard2014, Yang2016}. Moreover, the ensemble based approach offers excellent prospects for different multiplexing techniques \cite{Sangouard2011,Kutluer2017,Laplane2017,Pu2017}. The combination of the spin-wave to photon conversion capability together with the time-bin encoding makes our system a source of entangled photonic qubits that are robust and synchronizable. These capabilities provide an interesting resource for instance for long distance quantum communication using quantum repeaters \cite{Duan2001, Sangouard2011}.

\begin{figure}
	\includegraphics[width=.48\textwidth]{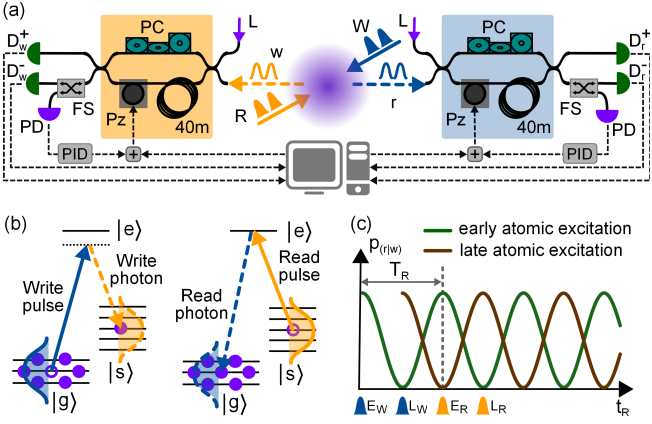}
	\caption{(color online) (a) Schematic overview of the experimental setup. W, write pulse; R, read pulse; w, write photon; r, read photon; L, interferometer lock light; PC, polarization controller; Pz, piezo-electric fiber stretcher; $\mathrm{D^{+(-)}_{w(r)}}$, single photon detectors; FS, fiber switch; PD, photodiode; PID, proportional-integral-derivative controller. (b) Energy levels relevant for the photon generation process. (c) Expected behaviour of the read photon transfer efficiency of the early (green) and late (brown) atomic excitations as a function of the read-out time $t_R$. The blue pulses indicate the times of the early ($\mathrm{E_W}$) and late ($\mathrm{L_W}$) write pulse peaks required to create the light-matter entangled state. The orange pulses indicate the times of the early and late read pulse peaks ($\mathrm{E_R}$ and $\mathrm{L_R}$) required to subsequently convert the atomic qubit into a read photon time-bin qubit.}
	\label{Figure1}
\end{figure}

The basic concept of the experiment is as follows (cf. Fig.~\ref{Figure1}). An off-resonant doubly-peaked \textit{write} laser pulse generates an excitation in our atomic cloud that is entangled with a Raman scattered write photon in the time-bin degree of freedom. In order to generate this entangled state, the atomic excitations generated at the two time bins need to form and orthogonal qubit basis (i.e. they need to be totally distinguishable). This is achieved applying an homogeneous magnetic field that leads to a Zeeman splitting of the atomic energy levels and induces a dephasing and rephasing of the atomic excitation at well defined times after its creation \cite{Matsukevich2006} (see Fig.~\ref{Figure1}(b) and \ref{Figure1}(c)). To assess the matter qubit, this subsequently converted into a photonic time-bin qubit using a resonant \textit{read} laser pulse, and the two entangled photons are analyzed. This analysis is done with Mach-Zehnder interferometers and single photon detectors, which allow projective measurements in any basis on the equator of the Bloch sphere. 

The experimental set-up is shown in Fig.~\ref{Figure1}(a). We cool an ensemble of $\mathrm{^{87}Rb}$ atoms in a magneto optical trap (MOT) to $T\approx100\,\mathrm{\mu K}$ \cite{Albrecht2015a}. The atomic levels relevant for the experiment are shown in Fig. \,\ref{Figure1}(b) and consist of two metastable ground states ($|g\rangle = |5^2S_{1/2},F=1\rangle$ and $|s\rangle = |5^2S_{1/2},F=2\rangle$) and one excited state ($|e\rangle = |5^2P_{3/2},F=2\rangle$). An homogeneous magnetic field ($B=2.1\,\mathrm{G}$) perpendicular to the beam propagation axis splits the energy of the Zeeman sub-levels, which is essential for the entanglement generation as discussed below. After optically pumping the atoms in state $|g\rangle$, a write pulse drives transition $|g\rangle\rightarrow|e\rangle$ with a red detuning of $\mathrm{\Delta=40MHz}$. This process generates probabilistically write photons on the $|e\rangle\rightarrow|s\rangle$ transition through spontaneous Raman scattering. The write pulse interacts with many atoms at different positions and in different Zeeman sublevels. Hence, the state of the atomic excitation is a collective state given by the superposition of all possible excitations \cite{Albrecht2015}. Due to its temporal evolution, the state can be written after a certain time $t$ as 
\begin{equation}
\left|\Psi_{\rm a}(t)\right\rangle=\frac{1}{\sqrt{N}}\sum_{j=1}^{N}e^{i(\bold{k}_{\rm W}-\bold{k}_{\rm w})\bold{x}_j+i\Delta w_jt}\left|g_1...s_j...g_N\right\rangle
\label{eqn:spinwave}
\end{equation}
where $N$ denotes the total number of atoms, $\bold{x}_j$ the initial atom position, $\Delta w_j$ the two photon detuning of the excitation and $\bold{k}_{\rm W(w)}$ the wavevector of the write pulse (photon).

\begin{figure}
	\includegraphics[width=.48\textwidth]{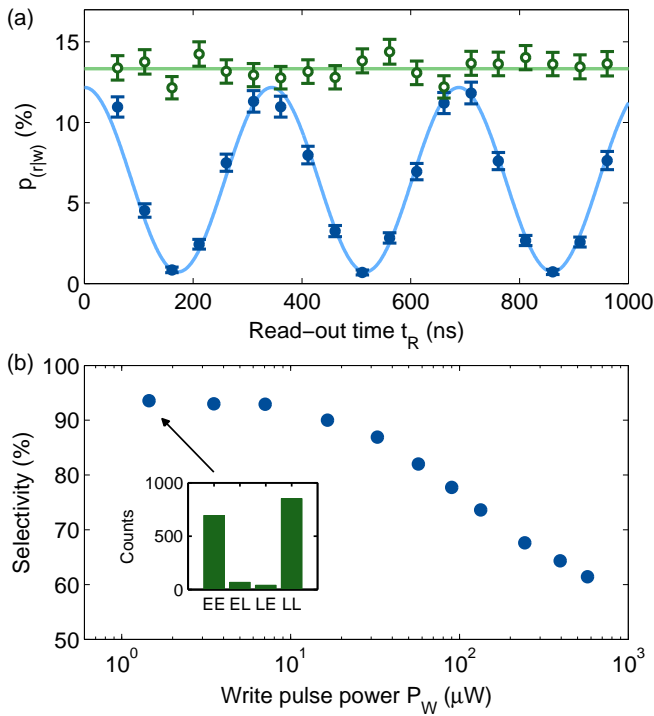}
	\caption{(color online) (a) Measured read photon transfer efficiency of the atomic excitation as a function of the read-out time. The green open circles show the case without magnetic field dephasing, while the blue dots show the case with the magnetic field on. For both traces the write (read) detection gates are $40\,\mathrm{ns}$ ($60\,\mathrm{ns}$) and the data are only corrected for the SPD detection efficiency. (b) Atomic excitation read-out selectivity as a function of the write pulse power. Here, the write and read detection gates were set to $30\,\mathrm{ns}$ and $40\,\mathrm{ns}$, respectively. The error bars are smaller than the size of the data points. The inset shows the time-bin correlation for one particular write pulse power. Error bars correspond to $\pm$1 standard deviations of the photon counting statistics.}
	\label{Figure2}
\end{figure}

The collective atomic excitation can be converted into a read photon by means of a read pulse resonant to the $|s\rangle\rightarrow|e\rangle$ transition. In the absence of atomic dephasing the transfer will happen with a high efficiency to a particular spatio-temporal mode thanks to collective interference of all contributing atoms. The spatial mode is given by the phase matching condition $\rm \bold{k}_r=\bold{k}_R+\bold{k}_W-\bold{k}_w$, where $\rm \bold{k}_{r(R)}$ are the read photon (pulse) wave vectors. This retrieval efficiency can be measured as the probability $p_{(r|w)}$ to detect a read photon once a write photon was detected, and is shown in Fig.~\ref{Figure2}(a) as a function of the read-out time $t_R$ if no external magnetic field is applied (green open circles). In the presence of dephasing of the atomic state, the collective interference will be degraded and the photon retrieval efficiency will decrease. In our case, the magnetic field splits the Zeeman sublevels, giving rise to four different excitation paths of the spin-wave with different two-photon detunings $\Delta\omega_j$ (cf. Supplemental Material \cite{supMat}). According to Eqn.~\eqref{eqn:spinwave}, this leads to a periodic de- and rephasing of the atomic excitation that we observe as a beating of the retrieval efficiency with a periodicity of $T_{\rm r}=344\,\mathrm{ns}$. This effect is schematically represented in Fig.\,\ref{Figure1}(c) and the measured data are shown in Fig.\,\ref{Figure2}(a) (blue dots). A theoretical model of $p_{\rm(r|w)}$ can be developed by computing the overlap between the initial atomic state and the state at the read-out time $p_{\rm(r|w)}(t_{\rm R})\propto\left|\left\langle\Psi_{\rm a}(t=0)|\Psi_{\rm a}(t=t_{\rm R})\right\rangle\right|^2$. This expression is used to obtain the equation with which the data in Fig.~\ref{Figure2}(a) are fitted \cite{supMat}.

The way we generate the entanglement with time-bin encoding is the following. A write pulse with two intensity peaks separated by $T_{\rm r}/2$ is sent to the atoms, leading to the probabilistic generation of a photon-atomic excitation pair delocalized in two time-bins. This time-bin entangled state can be written as
\begin{equation}
\left|\Psi_{\rm wa}\right\rangle=\frac{1}{\sqrt{2}}\left|\rm E_wE_a\right\rangle+e^{i\phi}\left|\rm L_wL_a\right\rangle
\end{equation}
where $\rm \left|E_{w(a)}\right\rangle$ and $\rm \left|L_{w(a)}\right\rangle$ denote a write photon (atomic excitation) generated in the early and late bin, respectively. The generation time difference of the atomic states $\rm \left|E_a\right\rangle$ and $\rm \left|L_a\right\rangle$ leads to a different phase evolution of these two spin-wave modes in such a way that they become distinguishable. The phase $\phi$ of the entangled state $\left|\Psi_{\rm wa}\right\rangle$ depends on the phase difference between the two write pulse peaks. 

Due to the mentioned collective interference, the atomic qubit can be later transferred into a photonic qubit via a resonant read pulse. This pulse needs to have two intensity peaks separated by $T_{\rm r}/2$ and delayed by a multiple of $T_{\rm r}$ with respect to the early atomic excitation creation time (cf. Fig.~\ref{Figure1}(c)). In that situation, thanks to the magnetic field dephasing, the early peak of the read pulse will transfer collectively only early atomic excitations into early read photons. In the same way, the late peak of the read pulse will  transfer collectively only late atomic excitations into late read photons. One detrimental effect is that the early read pulse also scatters late atomic excitations, which leads to the undirectional emission of photons. For this reason the early read pulse has an area of $\pi/2$, while the area of the late read pulse is $\pi$. This limits intrinsically the atom-to-photon state transfer to an efficiency of $50\%$ in the current configuration. As discussed in \cite{supMat}, this read-out imperfection is equivalent to a loss induced by a beam splitter. This issue could be overcome by an alternative scheme that transfers back and forth the early or late atomic excitations to different ground-state levels, such that during the write and read-out processes just one component of the spin-wave interacts with the optical pulses \cite{Jiang2016}.

To demonstrate that the generated write and read-photons are indeed correctly correlated in the time-bin degree of freedom, we construct the coincidence histograms of detection events, using just one write and one read photon detector after the ensemble. The inset in Fig.~\ref{Figure2}(b) shows the number of coincidences between write and read photons in each of the bins $C_{b_{\rm w}b_{\rm r}}$, where $b_{\rm w(r)}$ denotes the early or late write (read) photon bin. For low $P_\mathrm{W}$, most of the photons are detected together in either the early or the late time-bin. 
However, for higher write powers, the degree of correlation decreases, due to the creation of multiple spin-wave photon pairs. 
The green data points show the degree of correlation depending on the the write power --- illustrated by the selectivity parameter $S=\left(C_{\rm EE}+C_{\rm LL}\right)/\left(C_{\rm EE}+C_{\rm LL}+C_{\rm EL}+C_{\rm LE}\right)$. 

To characterize the generated quantum state and certify the entanglement, we took projective measurements on the equator of the Block sphere. This is done by overlapping the early and late bins of both write and read fields using fiber Mach-Zehnder interferometers as shown in Fig.~\ref{Figure1}(a). The length difference between the short and long arm of each interferometer is $40\,\mathrm{m}$, corresponding to the temporal separation between the early and late bins. This large imbalance makes it experimentally challenging to maintain a stable phase difference between the two arms of each interferometer during the whole experiment. In order to achieve that condition, both interferometers are actively temperature stabilized, and  a short section of the long fiber arm is rolled around a piezo-electric ceramic cylinder for active feedback. We send $325\,\mathrm{\mu W}$ of lock laser light, derived from the read pulse laser, to one input of each interferometer and measure the power at the output with a photodiode (PD). The PD signal is used to generate a feedback voltage which is sent to the piezo to keep the interferometer output power constant. This ensures that the phase difference of both interferometers is maintained at the required value during the experiment. The lock light and the write and read photons are not present at the same time: we perform repeatedly $13.3\,\mathrm{ms}$ of interferometer active stabilization during the atomic cloud preparation, followed by $1.4\,\mathrm{ms}$ of photon generation. In order to ensure polarization overlap at the interferometer output, we use polarization controllers in the short interferometer arms.
 
\begin{figure}
	\includegraphics[width=.48\textwidth]{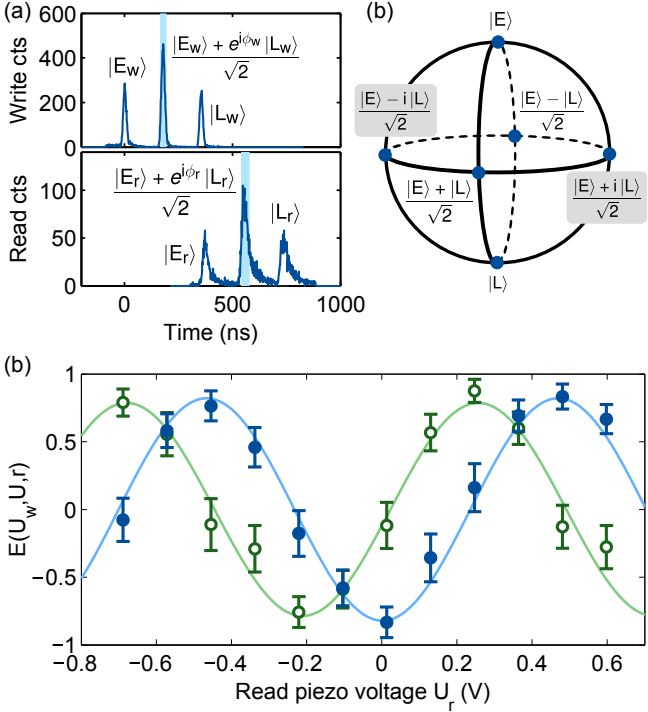}
	\caption{(color online) Write and read photon histograms at the interferometer outputs (photon durations are $20\,\mathrm{ns}$ and $30\,\mathrm{ns}$ FWHM, respectively). The detection events in each of the bins correspond to a projection to a different state as indicated in the figure. (b) Representation of the different orthogonal time-bin states on  the Bloch sphere. (c) Write-read photon correlation coefficients taken at $P_\mathrm{W}=7\,\mathrm{\mu W}$. The phase of the read photons interferometer is scanned, while the phase of the write photons interferometer is fixed at two different values ($U_{\rm w1}=0\,\mathrm{V}$ for the blue dots and $U_{\rm w2}=0.268\,\mathrm{V}$ for the green open circles). The detection gate widths were set to $30 \,\mathrm{ns}$ for the write photons and $40\,\mathrm{ns}$ for the read photons (cf. blue area in (a)).}
	\label{Figure3}
\end{figure}

Figure~\ref{Figure3}(a) shows the write and read photon time histograms after passing through the interferometers. For both histograms, the first peak corresponds to photons generated in the early bin that pass through the short interferometer arm. The third peak corresponds to photons generated in the late bin that pass through the long interferometer arm. The central peak corresponds to photons that are either created in the early bin and pass through the long interferometer arm or are created in the late bin and pass through the short interferometer arm. These events correspond to qubit projections into states that lay on the equator of the Bloch sphere expressed as $1/\sqrt{2}(\left|\rm E\right\rangle+e^{i\phi}\left|\rm L\right\rangle)$ \cite{Brendel1999,Marcikic2004} (see Fig.\,\ref{Figure3}(b)). The phase $\phi$ corresponds to the phase delay between the two interferometer arms, and can be controlled by changing the voltage $U_{\rm w(r)}$ sent to the write (read) piezo fiber stretcher. In Fig.\,\ref{Figure3}(c) we show the correlation coefficient $E$ as a function of $U_r$, defined as $E=(p_{++}-p_{+-}-p_{-+}+p_{--})/(\sum_{i,j}p_{i,j})$. Here, $p_{i,j}$ correspond to the probabilities to detect a photon coincidence between detectors $\mathrm{D}^i_\mathrm{w}$ and $\mathrm{D}^j_\mathrm{r}$ ($i,j=\pm$), within the central peaks shown in Fig.\,\ref{Figure3}(a). For two different $U_{\rm w}$, we measure two-photon interference fringes that are shifted by 82(5)$^\circ$ with visibilities of $V_1=0.82(0.04)$ and $V_2=0.79(0.04)$. These values reveal the presence of entanglement, since they are above the visibility required for a Bell inequality violation ($V>1/\sqrt{2}\approx0.707$). 

Besides the two-photon interference shown in Fig.~\ref{Figure3}(c), we also observe low visibility single-photon interference of each write and read photon \cite{supMat}. This is due to the emission of photons in other transitions that do not correspond to entangled events, which could be avoided by spectral filtering. However, these  uncorrelated write-read photons have a very low impact on $E(U_{\rm w},U_{\rm r})$.

\begin{figure}
	\includegraphics[width=.48\textwidth]{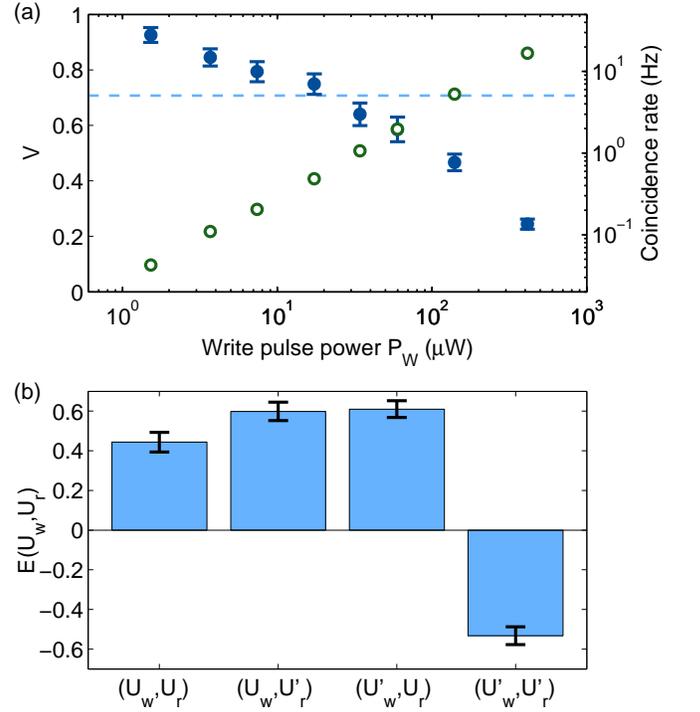}
	\caption{(color online)(a) Visibility of the two-photon interference fringes as shown in Fig.~\ref{Figure3}(c) (blue dots) and detected photon coincidence rate (green open circles) versus write pulse power. The dashed line represents the visibility required to violate a Bell inequality ($V>1/\sqrt{2}$). (b) Values of the four correlation coefficients taken at $P_W=3.5\mathrm{\mu W}$ and with the four basis settings ($U_{\rm w}=0.268\,\mathrm{V}$, $U_{\rm w}'=0\,\mathrm{V}$, $U_{\rm r}=0.333\,\mathrm{V}$, and $U_{\rm r}'=0.099\,\mathrm{V}$) that are optimal for a Bell inequality violation.}
	\label{Figure4}
\end{figure}

The visibility of the two-photon interference fringes (as shown in Fig.~\ref{Figure3}(c)) is plotted as a function of the write pulse power in Fig.~\ref{Figure4}(a). $V$ decreases for high $P_\mathrm{W}$ due to the creation of multiple photon-atomic excitation pairs during the write process \cite{Riedmatten2006}. This leads to the detection of coincidences coming from photon pairs that are not entangled. The measured data show that we can fulfill the condition of Bell inequality violation ($V>1/\sqrt{2}$) for $P_\mathrm{W}<15\,\mathrm{\mu W}$, reaching a maximum value of $V=0.93(5)$. To definitely prove the entanglement we violate a CHSH-type Bell inequality \cite{Clauser1969}, which reads
\begin{equation}
S=\left|E(\phi_{\rm w},\phi_{\rm r})+E(\phi_{\rm w},\phi_{\rm r}')+E(\phi_{\rm w}',\phi_{\rm r})-E(\phi_{\rm w}',\phi_{\rm r}')\right|\leq 2
\end{equation}
where $\phi_{\rm w(r)}$ and $\phi_{\rm w(r)}'$ are pairs of write (read) photon interferometer phases. In Fig.~\ref{Figure4}(b) we show the four correlation coefficients taken with the basis that lead to the highest possible value of $S$. The measurement gives a CHSH Bell parameter of $S=2.18\pm0.09$, violating the Bell inequality by 2 standard deviations. Possible limitations of this value include laser frequency fluctuations, the detection of photons  emitted in other transitions and the imperfect dephasing due to the photon temporal widths \cite{supMat}.

As can be observed in Fig.~\ref{Figure3}(a), the overlap of the early and late photonic modes in the interferometer is limited to a probability of 50$\%$. The events in the first and third peaks in the write and read photon histograms are discarded for our Bell inequality analysis. This limitation is due to the beam splitter at the input of the interferometers, and leads to a  Bell inequality loophole that is present in many of the photonic energy-time and time-bin entanglement experiments. In order to avoid it one could either replace the input beam splitter by an optical switch, or change the geometry of the experiment as described in \cite{Lima2010,Cuevas2013}.

In conclusion, we presented the direct generation of time-bin entanglement between a photon and a collective atomic spin excitation. After transferring the atomic qubit into a photonic state, the qubits are analyzed with Mach-Zehnder interferometers and we verify the entanglement by violating a CHSH Bell inequality. Photons with narrow linewidth as generated with our system \cite{Farrera2016a} are crucial for an optimal interaction with other narrowband matter quantum systems, like trapped ions, doped crystals, etc. Since photonic time-bin qubits are very suitable for quantum frequency conversion \cite{Tanzilli2005,Fernandez-Gonzalvo2013}, entanglement between hybrid matter systems could be achieved through the frequency conversion and distribution of the write photonic qubit \cite{Farrera2016,Ikuta2016,Maring2017}. 

\begin{acknowledgments}
We would like to thank M. Cristiani and B. Albrecht for their contributions at the early stage of the experiment. We acknowledge support by the ERC starting grant QuLIMA, by the Spanish Ministry of Economy and Competitiveness (MINECO) and the Fondo Europeo de Desarrollo Regional (FEDER) through grant FIS2015-69535-R, by MINECO Severo Ochoa through grant SEV-2015-0522, by Fundaci\'{o}  Privada Cellex and by the CERCA programme of the Generalitat de Catalunya. P.F. acknowledges the International PhD-fellowship program "la Caixa"-Severo Ochoa @ ICFO. G.H. acknowledges support by the ICFOnest+ international postdoctoral fellowship program.\\
\end{acknowledgments}

%\bibliography{DLCZtimeBinEntanglement}


\end{document}